\documentclass[twoside,english,1p,final,sort&compress]{elsarticle}
\usepackage[T1]{fontenc}
\usepackage[latin9]{inputenc}
\usepackage{amsthm}
\usepackage{gensymb}
\usepackage{xspace}
\usepackage{amsmath}
\usepackage{accents}
\usepackage{multirow}
\usepackage{xcolor}
\usepackage{amssymb}

\makeatletter

\theoremstyle{plain}

\journal{Finite Elements in Analysis and Design}

\makeatother

\usepackage{babel}
\providecommand{\theoremname}{Theorem}

\newcommand{\w}{\mbox{\large\ensuremath{\mathsf{w}}}}
\newcommand{\A}{\mbox{\large\ensuremath{\mathsf{A}}}}
\newcommand{\B}{\mbox{\large\ensuremath{\mathsf{B}}}}
\newcommand{\C}{\mbox{\large\ensuremath{\mathsf{C}}}}
\newcommand{\dotp}{\boldsymbol{\cdot}}
\newcommand{\ccirc}{\kern0.5ex\vcenter{\hbox{$\scriptstyle\circ$}}\kern0.5ex}
\newcommand{\e}[1]{\exp\left({#1}\right)}
\newcommand{\lay}[1]{^{(#1)}}
\newcommand{\mdot}[1]{\accentset{\mbox{\bfseries .}}{#1}}

\newcommand*{\versus}{\emph{vs.}\@\xspace}
\newcommand*{\eal}{et \emph{al.}\@\xspace}

\newcommand{\ERMS}{\text{E}_\text{RMS}}
\newcommand{\AARE}{\text{E}_\text{AAR}}

\DeclareMathOperator{\sigmoid}{sig}

\usepackage{esvect}
\def\traitfill@#1#2#3#4{%
  $\m@th\mkern2mu\relax#4#1\mkern-1.5mu %on met \relbaredd au d\'ebut
   \cleaders\hbox{$#4\mkern-0.3mu#2\mkern-0.3mu$}\hfill %remplit avec relbareda
   \mkern-1.5mu#3$%
}
\renewcommand{\overrightarrow}{\vv}

\begin{document}

\begin{frontmatter}{}

\title{Efficient Implementation of Non-linear Flow Law Using Neural Network into the Abaqus Explicit FEM code}

\author[LGP]{Olivier Pantalé \corref{cor1}}

\ead{Olivier.Pantale@enit.fr}
\ead[url]{http://www.enit.fr}

\author[LGP]{Pierre Tize Mha}

\author[LGP]{Amèvi Tongne}

\cortext[cor1]{Corresponding author}

\address[LGP]{Laboratoire Génie de Production, INP/ENIT, Université de Toulouse, 47 Av d'Azereix, Tarbes, France 65016}

\begin{abstract}
Machine learning techniques are increasingly used to predict material behavior in scientific applications and offer a significant advantage over conventional numerical methods. In this work, an Artificial Neural Network (ANN) model is used in a finite element formulation to define the flow law of a metallic material as a function of plastic strain $\varepsilon^p$, plastic strain rate $\mdot{\varepsilon}^p$ and temperature $T$. First, we present the general structure of the neural network, its operation and focus on the ability of the network to deduce, without prior learning, the derivatives of the flow law with respect to the model inputs. In order to validate the robustness and accuracy of the proposed model, we compare and analyze the performance of several network architectures with respect to the analytical formulation of a Johnson-Cook behavior law for a 42CrMo4 steel. In a second part, after having selected an Artificial Neural Network architecture with $2$ hidden layers, we present the implementation of this model in the Abaqus Explicit computational code in the form of a VUHARD subroutine. The predictive capability of the proposed model is then demonstrated during the numerical simulation of two test cases: the necking of a circular bar and a Taylor impact test. The results obtained show a very high capability of the ANN to replace the analytical formulation of a Johnson-Cook behavior law in a finite element code, while remaining competitive in terms of numerical simulation time compared to a classical approach.

\end{abstract}

\end{frontmatter}{}

\section{Introduction\label{sec-1:I}}

Numerical simulation of forming processes, machining or the behavior of structures subjected to dynamic loads and impacts requires the use of specific material behavior laws, whose parameters are identified by tests based on Taylor impacts, Hopkinson bars or Gleeble thermomechanical simulator. The behavior laws are selected according to their availability in a finite element code or the possibility, if not available, to implement them through user subroutines. In this study, our work is based on the use of the finite element code Abaqus Explicit which offers the possibility to define user behavior laws through FORTRAN subroutines VUMAT or VUHARD \cite{Gao-2007-FRT, JansenVanRensburg-2012-TSV} like the work proposed by Duc-Toan \eal \cite{Duc-Toan-2012-MJC} or more recently by Ming \eal \cite{Ming-2018-ERV}. The usual procedure is to select a mathematical form of the behavior law among those available in the literature (Johnson-Cook, Zerilli Armstrong, ...) and then, from the results of experimental tests, to identify via a regression method, the parameters of the selected law. 

In most cases, the behavior of the material at high temperatures and strain rates is highly nonlinear, and the effects of many factors on the flow stress are also nonlinear, which reduces the accuracy of the prediction by the regression methods usually used and limits the field of application. In addition, the selection, development, and numerical implementation of such constitutive equations is time-consuming. Artificial intelligence techniques allow advances concerning the laws of behavior in order to allow a better identification of these laws. Thus, Versino \eal \cite{Versino-2017-DDM} used a Machine Learning technique based on symbolic regression for the development of data-driven constitutive model. Obtaining a flow equation, and thus its analytical derivative, allows the use of iterative solvers that employ Jacobians (i.e., the Newton-Raphson scheme) that allow a higher order of convergence. This symbolic regression technique has also been used by other authors since, such as Bomarito \eal \cite{Bomarito-2021-DID}, Park \eal \cite{Park-2021-MCM} using constrained symbolic regression technique, or Nassr \eal \cite{Nassr-2018-DCM} using evolutionary polynomial regression.

Given this situation, it is therefore natural to look for a method to eliminate some intermediate steps between experimental tests and numerical simulation in order to simplify the computational chain. In this perspective, recent advances in deep learning constitute a way of investigation.
The basic idea is to replace the analytical formulation used to calculate the flow stress $\sigma$ of the material as a function of the plastic strain $\varepsilon^p$, the plastic strain rate $\mdot{\varepsilon}^p$ and the temperature $T$, by an Artificial Neural Network (ANN). This neural network is trained to reproduce the behavior of the considered material only from the experimental data resulting from the tests, ignoring any assumption on the analytical form of the assumed flow law. Consequently, it is no longer necessary to postulate an analytical form of the behavior law in order to implement it in a FEM code. 

Artificial neural networks and deep learning are becoming more and more important in today's society, and their fields of application are getting wider and wider. After a boom in the early 1990s and a decline in interest towards the end of the $20^{th}$ century, neural networks are experiencing a resurgence of interest and even a huge media hype under the name of deep learning. Their use in science and physics is now widespread, notably because of the current availability of efficient tools allowing to program Artificial Neural Networks thanks to widely available libraries such as Tensorflow \cite{Mattmann-2020-MLT} for example. The most publicized applications of deep learning are mainly related to medical diagnosis, robotics, images and language recognition, but the applications go far beyond that and all sciences can now use these techniques, including thermomechanical numerical simulation.

Artificial neural networks can solve problems that are difficult to conceptualize using traditional computational methods. Unlike a classical approach based on a regression method, an Artificial Neural Network does not need to know the mathematical form of the model it seeks to reproduce. The Artificial Neural Network learns from the training data and can reproduce the behavior of a model from the simple knowledge of a series of input and output values with no prior assumption on their nature and their interrelations.
Hornik \eal \cite{Hornik-1989-MFN} have rigorously established in 1989 that feed-forward Neural Network are a class of universal approximators, extending the work proposed twenty years before by Minsky and Papert \cite{Minsky-1969-PIC} where they demonstrated that the simple two-layer perceptron is incapable of usefully representing or approximating functions outside a very narrow and special class.
The ANN has adjustment, memorization and anticipation capabilities, and better performances than the approach based on implementing a constitutive equation. Therefore, Artificial Neural Networks can now enable novel approaches for modeling the behavior of materials and have been successfully applied in the prediction of constitutive relationships of some metals and alloys in recent years. Applicability of ANN to model path dependent plasticity has been explored and some review of the literature can be found for example in Gorji \eal \cite{Gorji-2020-PRN} concerning the use of Recurrent Neural Network, in Jamli \eal \cite{Jamli-2019-SNN} concerning their application in finite element analysis of metal forming processes, or in Jiao \eal \cite{Jiao-2020-AIS} concerning the applicability to meta-materials and their characterization. A distinction must be made between ANN-based hardening models and ANN-based constitutive models. Both approaches have been studied by many researchers over the last thirty years. Ghaboussi \eal \cite{Ghaboussi-1991-KBM} published a pioneering paper, in which they proposed an ANN-based constitutive model for planar concrete under monotonic biaxial loading and cyclic uniaxial loading in which they successfully predicted several loading paths in the biaxial loading condition. They improved NN architecture by introducing Adaptive NN and Autoprogressive NN in \cite{Ghaboussi-1998-ATN, Ghaboussi-1998-NNA} where the network architecture evolves during the training phase to better learn complex stress-strain behavior of materials using a global load-deflection response. The approach adopted for this study is an ANN-based hardening model for which, the evaluation of the material flow stress calculated by the ANN is combined with a Radial Return type integration scheme.

Lin \eal \cite{Lin-2008-ANN} developed a neural network to predict the flow stress of 42CrMo4 steel in hot compression tests on a Gleeble thermomechanical device and showed a very good correlation between experimental and predicted results. An extension to the numerical implementation of this approach would have been desirable.
Javadi \eal \cite{Javadi-2009-IFE} used a neural network to capture the behavior of complex materials using a FE model incorporating a backpropagation neural network.
Lu \eal \cite{Lu-2011-ANN} presented a comparative study of the modeling of an Al-Cu-Mg-Ag alloy behavior by a constitutive equation based on the Zener-Hollomon parameter and a neural network. It also shows that the model based on the ANN proposes a better prediction than the constitutive equation.
Ashtiani \eal \cite{Ashtiani-2016-CSP} compared the prediction capabilities of an ANN against a conventional approach based on several behavior laws such as Johnson-Cook, Arrhenius and Strain compensated Arrhenius and concluded better efficiency and accuracy of the neural network in predicting the hot behavior of Al-Cu-Mg-Pb alloy. Ashtiani \eal \cite{Ashtiani-2016-CSP} have shown that a well-trained ANN can efficiently overcome the lacks of physics coming from analytical constitutive behaviors such as the Johnson-Cook, or the Arrhenius one.
Ali \eal \cite{Ali-2019-AAN} presented an ANN model coupled with a rate dependent crystal plasticity finite element method to simulate the stress-strain behavior of material and its microstructure evolution in a AA6063-T6 under simple shear and tension.
Stoffel \eal \cite{Stoffel-2018-ANN, Stoffel-2019-NNB} applied ANN to complicated structural deformations of shock-wave loaded plates involving both geometrical and physical non-linearities.
Li \eal \cite{Li-2019-MBT} implemented a VUMAT subroutine for Abaqus where parameters were identified through a combination of analytical formulas and a back propagation algorithm.
Recently, Huang \eal \cite{Huang-2021-CMM} developed a neural network model to predict the flow stress and the microstructure evolution of Ti-6Al-4V alloy. It also showed the superiority of this approach over an Arrhenius behavior model, especially because the ANN can predict the flow stress in the whole range of deformation.
Temporal Convolutional Networks have also been applied by Abueidda \eal \cite{Abueidda-2021-DLP} to predict the history-dependent responses of a class of cellular materials. Thermodynamics based ANN where also proposed by Masi \eal \cite{Masi-2021-TAN} to reduce physically inconsistencies in the predictions of the NN. They demonstrate the wide applicability of TANNs for modeling elasto-plastic materials, using
both hyper- and hypo-plasticity models.
As presented by Knight \eal \cite{Knight-2021-LGB}, evolution of Neural Network Architectures is not the only way to progress, constant evolution of the hardware architecture to implement ANN will also have to be taken into account for the next future.

In section \ref{sec-2:ANN-Setup}, we present the main bases of the deep learning with in details the description of the structure of the neural network and the equations which govern its functioning. As we will see in section \ref{sec-4:ImpNNAbaExpl} concerning the numerical implementation of the flow law, the programming of the neural network on the finite element code Abaqus Explicit requires the determination of the $3$ derivatives of the flow stress $\sigma$ with respect to the plastic strain $\varepsilon^p$, the plastic strain rate $\mdot{\varepsilon}^p$ and the temperature $T$. The determination of these derivatives will be the subject of the second part of section \ref{sec-2:ANN-Setup}. Section \ref{sec-3:TrainPerfEval} is devoted to a detailed presentation of the ANN learning for a Johnson-Cook type behavior law for a 42CrMo4 steel. In this section, we will show the influence of the network structure on the accuracy of the flow stress and the derivatives evaluation.
Section \ref{sec-4:ImpNNAbaExpl} is devoted to a presentation of the numerical implementation of the neural network in the Abaqus finite element code in the form of a FORTRAN VUHARD subroutine as well as numerical test cases to validate the proposed approach. A conclusion and perspectives are finally proposed at the end.

\section{Artificial Neural Network set-up\label{sec-2:ANN-Setup}}

In this section, we briefly introduce the basic concepts of backward and forward propagating Artificial Neural Networks (ANNs) that are relevant to this work. The global architecture we have retained for this work is a multi-layer feed-forward network which can be seen as a universal approximator as proposed by Hornik \eal \cite{Hornik-1989-MFN}. The proposed neural network is used to approximate non-linear functions. The concept of neural network consists in simulating the flow of information inside the human brain by defining a set of neurons (arranged in different layers) defining functions producing results according to the inputs. These neurons are interconnected from one layer to another and continuously improve their predictive ability as the network is trained and learns a new concept \cite{Specht-1991-GRN}.
Figure \ref{fig:ANN-multilayer} presents the global architecture of an ANN with multiple hidden layers where neurons are set up in different layers (from the first hidden layer to the output layer) with the following characteristics:
\begin{figure}[h] 
  \centering
  \includegraphics[width=\columnwidth]{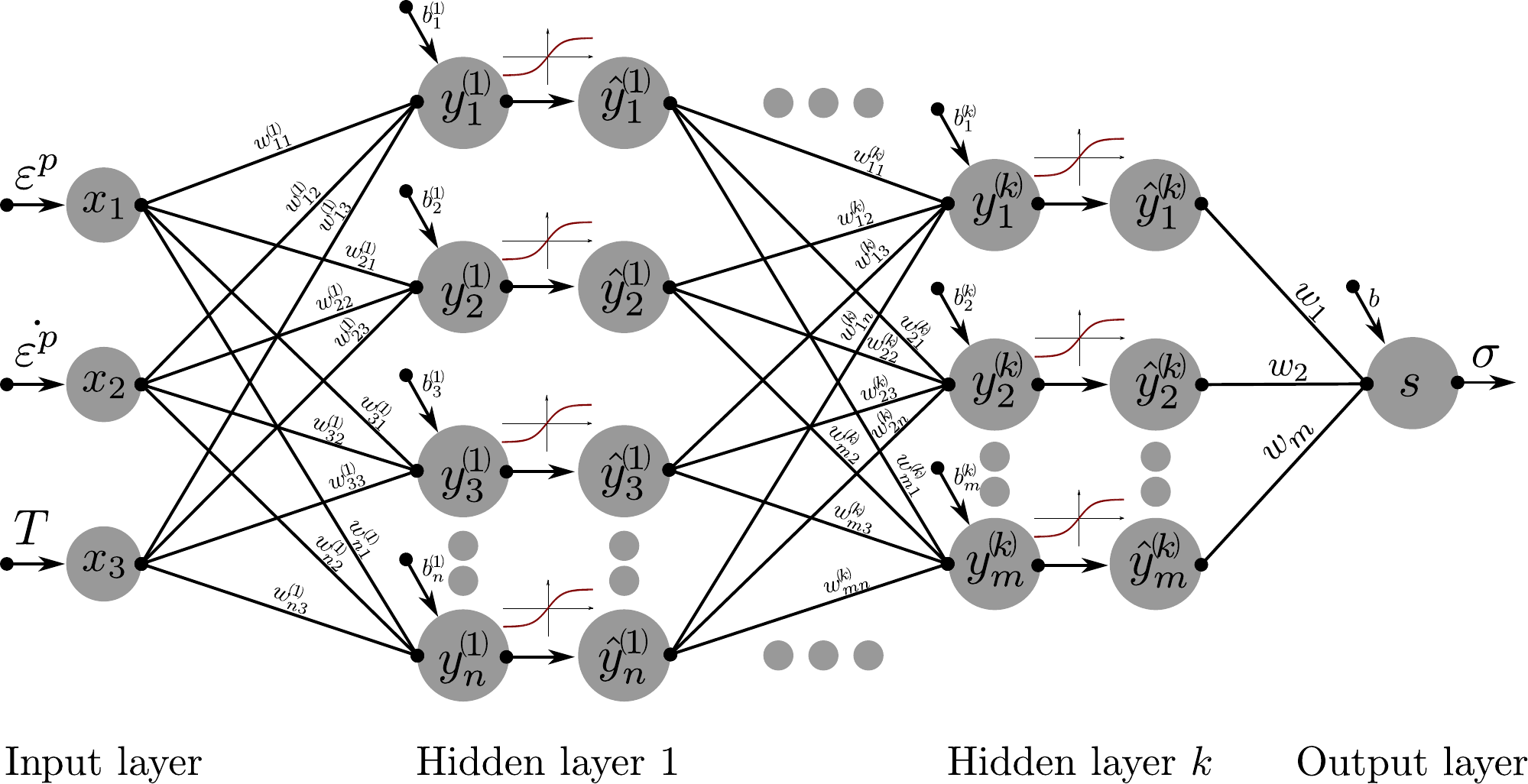}
  \caption{Multi-layer Artificial Neural Network architecture}
  \label{fig:ANN-multilayer}
\end{figure}
\begin{itemize}
\item The first layer, is the so-called input layer, and in our application concerning constitutive law, consists of 3 inputs. This one is not composed of neurons as they will be defined hereafter, and collects the incoming information: the three values corresponding to the plastic strain $\varepsilon^p$, the plastic strain rate $\mdot{\varepsilon}^p$ and the temperature $T$ respectively.
\item The ANN contains at least one hidden layer (in the current paper, we will consider only one and two hidden layers ANNs) containing a variable number of neurons.
\item The last layer is the so-called output layer, and in our case consists of only one neuron providing the value of the von Mises flow stress $\sigma$.
\item Neurons are not connected to other neurons of the same layer but only to the immediately preceding outputs and following inputs.
\end{itemize}
In Figure \ref{fig:ANN-multilayer} we have separated the summation $\overrightarrow{y}\lay{k}$ and the activation $\overrightarrow{\hat{y}}\lay{k}$ parts of all neurons within the hidden layers. Concerning the notation used, note here that we are using the formalism of a vector (using an arrow over the symbol) for many quantities in the equations hereafter, even if these are not vectors in the sense of a mathematical notation. In reality, these quantities are vertical one-dimensional arrays used to store data, which visually correspond to the components of a vector. In the proposed architecture, there is no activation function associated to the neuron in the output layer, as usually done for regression problems which is our case here. The data flows from layer to layer until we obtain the output of the ANN.
Every layer, except the first one, in the ANN, no matter how many neurons $n$ are in, have an input with a variable number of items $m$ and an output with a fixed number of items $n$. In the so-called forward propagation algorithm, the output of one layer becomes the input of the next one.

\subsection{Neural Network governing equations\label{subsec-2:NNGovEqu}}

\subsubsection{Input layer\label{subsec-2:InpLay}}
Conforming to Figure \ref{fig:ANN-multilayer}, the input layer is defined by $\overrightarrow{x} = [x_1, x_2, x_3]^T$. In the proposed application, $\overrightarrow{x}$ has three components $x_i$ related to the plastic strain $\varepsilon^p$, the plastic strain rate $\mdot{\varepsilon}^p$ and the temperature $T$, respectively.
As it will be presented in section \ref{subsec-2:PrePostData}, input variables $x_i$ are normalized to avoid ill-conditionning for the optimization procedure, mainly because they represent various physics with a large discrepancy in values.

\subsubsection{Hidden layers\label{subsec-2:HidLay}}
Any hidden layer $k$, containing $n$ neurons, takes a weighted sum of the outputs $\overrightarrow{x}$ of the immediately previous layer $(k-1)$, containing $m$ neurons, given by the following equation:
\begin{equation}
y_i\lay{k} = \sum_{j=1}^m w_{ij}\lay{k} x_j+ b_i\lay{k}\label{eq:ykind}
\end{equation}

where $y_i\lay{k}$ is the nodal value of the $i^{th}$ neuron of layer $k$, $w_{ij}\lay{k}$ is the associated weight parameter between the $i^{th}$ neuron of layer $k$ and the $j^{th}$ neuron of layer $(k-1)$ and $b_i\lay{k}$ is the associated bias of the $i^{th}$ neuron of layer $k$. Those weights and bias are the training parameters of the ANN that we have to adjust during the training procedure described in section \ref{sec-3:TrainPerfEval}. Using matrix notation, equation (\ref{eq:ykind}) can be rewritten with the following form:
\begin{equation}
\overrightarrow{y}\lay{k} = \w\lay{k} \dotp \overrightarrow{x}+ \overrightarrow{b}\lay{k}\label{eq:yk}
\end{equation}
where $\overrightarrow{y}\lay{k} = [y_1\lay{k}, y_2\lay{k},\cdots, y_n\lay{k}]^T$ contains the nodal values resulting from the summation operation in layer $k$, $\overrightarrow{b}\lay{k} = [b_1\lay{k}, b_2\lay{k},\cdots, b_n\lay{k}]^T$ is the nodal bias of layer $k$ and $\w\lay{k}$ is the $[n\times m]$ weight parameters matrix of layer $k$ given hereafter:
\begin{equation}
{\w\lay{k}}=\left[
\begin{array}{cccc}
w_{11}\lay{k} & w_{12}\lay{k} & \cdots & w_{1m}\lay{k}\\
w_{21}\lay{k} & w_{22}\lay{k} & \cdots & w_{2m}\lay{k}\\
\vdots & \vdots & \ddots & \vdots\\
w_{n1}\lay{k} & w_{n2}\lay{k} & \cdots & w_{nm}\lay{k}
\end{array}
\right]
\end{equation}
The total number of training parameters $N$ for any hidden layer $k$ is the sum of the number of weight parameters and the number of bias parameters of layer $k$, so $N=n(m+1)$.
After the summation operation defined by equation (\ref{eq:yk}), each neuron in the hidden layer $k$ provides an output value $\overrightarrow{\hat{y}}\lay{k}$ computed from an activation function $f\lay{k}$ according to the following equation:
\begin{equation}
\hat{y}_i\lay{k}=f\lay{k}\left(y_i\lay{k}\right)\qquad \text{or}\qquad\overrightarrow{\hat{y}}\lay{k}=f\lay{k}\left(\overrightarrow{y}\lay{k}\right)
\end{equation}
Many activation functions are available in literature and their choice depend mainly on the application of the ANN. In our case, it is very important that those activation functions are derivable in order to allow the computation of the derivative of the von Mises stress $\sigma$ with regard to the plastic strain $\varepsilon^p$, the plastic strain rate $\mdot{\varepsilon}^p$ and the temperature $T$. For our type of application, we have made the choice to test two among the mainly used ones: 
\begin{itemize}
\item the Sigmoid $\sigmoid(x)$ activation function defined by:
\begin{equation}
\sigmoid(x)=\frac{1}{1 + \e{-x}}\ ,\quad \sigmoid'(x)=\frac{\e{x}}{\left(1 + \e{x}\right)^2} \label{eq:sigmoid}
\end{equation}
\item and the Hyperbolic tangent $\tanh(x)$ activation function defined by:
\begin{equation}
\tanh(x)=\frac{\e{x}-\e{-x}}{\e{x}+\e{-x}}\ ,\quad \tanh'(x) = 1 - \tanh^2(x)\label{eq:tanh}
\end{equation}
\end{itemize}

The main guide for this choice is that the expression of their derivatives is simple, which leads to relatively compact expressions. Comparison and performance of those two activation functions will be presented in section \ref{sec-3:TrainPerfEval}.

If there exists another hidden layer $(k+1)$ after the current layer, then, the output $\overrightarrow{\hat{y}}$ of layer $k$ is then used as the input $\overrightarrow{x}$ of layer $(k+1)$.

\subsubsection{Output layer\label{subsec-2:OutLay}}
The output $s$ is computed from the values of the last hidden layer $l$ of the neural network, containing $m$ neurons, using the following equation:
\begin{equation}
s = \sum_{j=1}^m w_j \hat{y}_j\lay{l} + b
\end{equation}
where $b$ is the bias associated to the output neuron and $w_i$ are the $m$ weight parameters between the last hidden layer and the output neuron $s$. Using matrix formalism, one can rewrite this later as:
\begin{equation}
s = \overrightarrow{w}^T \dotp \overrightarrow{\hat{y}}\lay{l} + b\label{eq:ANN-output}
\end{equation}
with $\overrightarrow{w}=[w_1, w_2,\cdots,w_m]^T$. As presented earlier, there is no activation function for the output neuron, so $s$ is directly the output of the neural network. The total number of training parameters for the output layer is $m+1$. For a neural network with two hidden layers having $m$ neurons on the first hidden layer and $n$ neurons on the second one, the total number of training parameters is $N=4m+n(m+2)+1$.

\subsubsection{Pre and postprocessing of data\label{subsec-2:PrePostData}}
Since the ANN are set up to treat values with a limited amplitude, it is necessary to pre-process the provided corresponding values of the plastic strain $\varepsilon^p$, the plastic strain rate $\mdot{\varepsilon}^p$ and the temperature $T$ in the range $[0,1]$ in a same manner as the one proposed by other authors \cite{Lin-2008-ANN, Lu-2011-ANN}. Therefore, the input $\overrightarrow{x}$ of the ANN is computed according to the constitutive flow law by:
\begin{itemize}
\item Since the plastic deformation rate in constitutive equations usually enhances the logarithm of the plastic strain rate, we preprocess the plastic strain rate by computing the natural logarithm of the ratio of the plastic strain rate $\mdot{\varepsilon}^p$ over the referential strain rate $\mdot{\varepsilon_0}$ given by $\ln (\mdot{\varepsilon}^p/\mdot{\varepsilon_0})$.
\item Then, we normalize the $x_i$ variables in the range $[0,1]$ to avoid ill-conditioned system as presented by many other authors in the literature \cite{Lin-2008-ANN, Lu-2011-ANN}.
\end{itemize}
Therefore, the three components of the input $\overrightarrow{x}$ are obtained from the plastic strain $\varepsilon^p$, the plastic strain rate $\mdot{\varepsilon}^p$ and the temperature $T$ using the following expressions:
\begin{equation}
\overrightarrow{x} =
\begin{cases}
x_1 = \frac{\varepsilon^p - [\varepsilon^p]_{min}}{[\varepsilon^p]_{max} - [\varepsilon^p]_{min}}\\
x_2 = \frac{\ln(\mdot{\varepsilon}^p/\mdot{\varepsilon_0})-[\ln(\mdot{\varepsilon}^p/\mdot{\varepsilon_0})]_{min}}{[\ln(\mdot{\varepsilon}^p/\mdot{\varepsilon_0})]_{max}-[\ln(\mdot{\varepsilon}^p/\mdot{\varepsilon_0})]_{min}}\label{eq:preprocess}\\
x_3 = \frac{T-[T]_{min}}{[T]_{max}-[T]_{min}}
\end{cases}
\end{equation}
where $[\ ]_{min}$ and $[\ ]_{max}$  are the boundaries of the range of the corresponding field.
During the training of the ANN, the von Mises stresses $\sigma$ will also be scaled down within the range $[0,1]$, using the following expression:
\begin{equation}
s =  \frac{\sigma-[\sigma]_{min}}{[\sigma]_{max}-[\sigma]_{min}}\label{eq:postprocess}
\end{equation}
So, finally, the von Mises stress $\sigma$ can be obtained from the output $s$ of the ANN using:
\begin{equation}
\sigma =  \left([\sigma]_{max}-[\sigma]_{min}\right)s + [\sigma]_{min} \label{eq:vonMises-ANN}
\end{equation}
The $[\ ]_{min}$ and $[\ ]_{max}$ values of plastic strain, plastic strain rate, temperature, stresses and referential strain rate $\mdot{\varepsilon_0}$ used during the training phase should be recorded for later use during the implementation of the ANN in the Abaqus Explicit code. They are part of the final solution along with the weights $\w\lay{k}$ and bias values $\overrightarrow{b}\lay{k}$ of the hidden layers and the weights $\overrightarrow{w}$ and the bias $b$ of the output layer of the neural network that constitutes the training parameters of the neural network. The knowledge of these quantities after the learning phase allows then to extract the whole neural network from its formulation in Python language to a compact version, without the back propagation learning mechanism, for its implementation in FORTRAN in the Abaqus Explicit FEM code.

\subsubsection{Loss function\label{subsec-2:LossFunc}}
In the context of optimization algorithms, the function used to evaluate the quality of a solution is called the objective function. In the application to neural networks, we try to minimize the error made by the network in the prediction of the solution. The evaluation of this error consists in measuring the difference between the predicted value $\sigma_i$ computed by the NN and a reference value $\sigma_i^y$ coming usually from experiments, or, in the present paper from an analytical equation, that we want to reach for a particular data set.
There are several ways to define this error, among them, the best known is probably the average Root Mean Square Error ($\ERMS$) given by:
\begin{equation}
\ERMS = \sqrt{\frac{1}{N}\sum_{i=1}^{N}{(\sigma_i-\sigma^y_i)^2}}\label{eq:RMSE}
\end{equation}
where $N$ is the total number of values of the training batch.

\subsection{Derivatives computation\label{subsec:DC}}
Implementation of the neural network within the radial return algorithm proposed by Ming \eal \cite{Ming-2018-ERV} through a VUMAT subroutine of directly through a VUHARD one for Abaqus Explicit requires that the ANN returns not only the von Mises equivalent stress $\sigma$ given by equation (\ref{eq:vonMises-ANN}), but also the three derivatives $\partial \sigma/\partial \varepsilon^p$, $\partial \sigma/\partial\mdot{\varepsilon}^p$ and $\partial \sigma/\partial T$ without having been trained to compute those derivatives. So that we do not have $4$ outputs for the ANN but only one and the ability to compute those derivatives must be intrinsic. Therefore, we need to be able to compute those three derivatives using only the proposed ANN architecture. One straightforward, but not recommended, solution to this problem is to numerically compute the derivative of $\sigma$ with respect to $\varepsilon^p$, $\mdot{\varepsilon}^p$ and $T$ using the following relation:
\begin{equation}
\frac{\partial \sigma(x)}{\partial x} = \frac{\sigma(x+\delta x) - \sigma(x)}{\delta x}
\end{equation}
where $\delta x$ is a small increase ($\delta x=10^{-6}$ for example) applied to one of the $3$ variables $\varepsilon^p$, $\mdot{\varepsilon}^p$ and $T$. We need to compute $4$ times a result from the ANN to compute the flow stress and approximate the three derivatives which is quite time-consuming.

Another way to compute those derivatives is to compute the quantity $\overrightarrow{s}'=[s'_1,s'_2,s'_3]^T$ containing the $3$ derivatives of the output $s$ defined by equation (\ref{eq:ANN-output}) of the ANN with respect to the input $\overrightarrow{x}$. This analytic computation depends on the number of hidden layers (hereafter we present the results for $1$ and $2$ hidden layers), and the type of activation functions used (the $\tanh$ and the $\sigmoid$ functions). $\overrightarrow{s}'=\partial s/\partial \overrightarrow{x}$ contains the derivatives of $s$ with respect to $x_1$, $x_2$ and $x_3$ respectively. Based on the chain-rule derivation, with $1$ hidden layer and a $\tanh$ activation function, $\overrightarrow{s}'$ takes the following form:
\begin{equation}
\overrightarrow{s}' =
{\w\lay{1}}^T \dotp \left[ \overrightarrow{w} - \overrightarrow{w} \ccirc \tanh^2\left(\overrightarrow{y}\lay{1}\right)\right]\label{eq:der-1-tanh}
\end{equation}
where $\overrightarrow{y}\lay{1}$ is given by equation (\ref{eq:yk}) with $k=1$ and  $\ccirc$ is the element-wise product, known as the Hadamard product, which is a binary operation that takes two matrices $\A$ and $\B$ of the same dimensions and produces another matrix $\C$ of the same dimension as the operands, where each element $C_i=A_i~B_i$. The $\tanh$ operator applied on quantity $\overrightarrow{y}\lay{1}$ is just computed for each component of $\overrightarrow{y}\lay{1}$, as presented in Figure \ref{fig:ANN-multilayer}, since this is not a real but only an array of reals stacked vertically. With $1$ hidden layer and a $\sigmoid$ activation function, we obtain:
\begin{equation}
\overrightarrow{s}' =
{\w\lay{1}}^T \dotp \left[ \frac{\overrightarrow{w} \ccirc \e{-\overrightarrow{y}\lay{1}}}
{\left[1 + \e{-\overrightarrow{y}\lay{1}}\right]^2}
\right]\label{eq:der-1-sig}
\end{equation}

When the number of hidden layers increases, the complexity of the derivative increases, so, for $2$ hidden layers and a $\tanh$ activation function for both layers we obtain:
\begin{equation}
\overrightarrow{s}' =
{\w\lay{1}}^T \dotp \left[{\w\lay{2}}^T \dotp \left(\overrightarrow{w} - \overrightarrow{w} \ccirc \tanh^2\left(\overrightarrow{y}\lay{2}\right)\right)
\ccirc \left(1 - \tanh^2\left(\overrightarrow{y}\lay{1}\right)\right)\right]\label{eq:der-2-tanh}
\end{equation}

Finally, for $2$ hidden layers and a $\sigmoid$ activation function for both layers we obtain:
\begin{equation}
\overrightarrow{s}' =
{\w\lay{1}}^T \dotp \left[{\w\lay{2}}^T \dotp \left( 
\frac{\overrightarrow{w} \ccirc \e{-\overrightarrow{y}\lay{2}}}
{\left[1 + \e{-\overrightarrow{y}\lay{2}}\right]^2}\right)\ccirc \left(\frac{
\e{-\overrightarrow{y}\lay{1}}}
{\left[1 + \e{-\overrightarrow{y}\lay{1}}\right]^2}
\right)\right]\label{eq:der-2-sig}
\end{equation}

Depending on the number of hidden layers and the type of activation functions used, from equations (\ref{eq:der-1-tanh}) to (\ref{eq:der-2-sig}) and because of the pre and post-processing of the quantities $\sigma$, $\varepsilon^p$, $\mdot{\varepsilon}^p$ and $T$ defined in section \ref{subsec-2:PrePostData} by equations (\ref{eq:preprocess}) and (\ref{eq:postprocess}) one can finally obtain the derivatives of the flow stress $\sigma$ with respect to $\varepsilon^p$, $\mdot{\varepsilon}^p$ and $T$ using the following expression:

\begin{equation}
\begin{cases}
\partial \sigma/\partial \varepsilon^p = s'_1 \frac{[\sigma]_{max} -[\sigma]_{min}}{[\varepsilon^p]_{max} -[\varepsilon^p]_{min}}\\
\partial \sigma/\partial\mdot{\varepsilon}^p = \frac{s'_2}{\mdot{\varepsilon}^p} \frac{[\sigma]_{max} -[\sigma]_{min}}{[\mdot{\varepsilon}^p]_{max} -[\mdot{\varepsilon}^p]_{min}} \\
\partial \sigma/\partial T = s'_3 \frac{[\sigma]_{max} -[\sigma]_{min}}{[T]_{max} -[T]_{min}}
\end{cases}\label{eq:derivatives-ANN}
\end{equation}

It is important to note here, that whatever the method adopted to calculate the derivative of the output of the neural network with respect to the inputs (numerical method or formulation of the derivative of the network), the result obtained is an approximation of the derivative of the mathematical function represented by the network with respect to the inputs. Indeed, the neural network, by its construction, approximates a mathematical formulation, but, as shown by Nguyen \eal \cite{Nguyen-Thien-1999-AFT}, a neural network can also approximate the derivatives of the mathematical function it reproduces. Thus, the derivative of the neural network with respect to its inputs is correlated to the derivative of the mapped mathematical function with respect to its parameters.

\section{Training of the ANN and performance evaluation\label{sec-3:TrainPerfEval}}
In order to evaluate the performance of the proposed approach, we decided to reproduce the behavior of the Johnson-Cook \cite{Johnson-1983-CMD} flow law with an Artificial Neural Network because it is one of the most widely used flow law for the simulation of high strain rate deformation processes. It is implemented in numerous Finite Element codes such as Abaqus. Reproducing the behavior of a Johnson-Cook law by a neural network is of course not an aim of this work, but only required in order to verify numerically that the neural network is able to take into account a non-linear behavior and to have a way to measure exactly the prediction errors of the neural network. The general formulation $\sigma^{y}(\varepsilon^p,\mdot{\varepsilon}^p,T)$ is given by the following equation:
\begin{equation}
\sigma^{y}=\left(A+B\varepsilon^{p^{n}}\right) \left[1+C\ln\left(\frac{\mdot{\varepsilon}^p}{\mdot{\varepsilon}_0}\right)\right] \left[1-\left(\frac{T-T_0}{T_m-T_0}\right)^{m}\right]\label{eq:Johnson-Cook}
\end{equation}
where $\mdot{\varepsilon}_0$ is the reference strain rate, $T_0$ and $T_m$ are the reference temperature and the melting temperature of the material respectively and $A$, $B$, $C$, $n$ and $m$ are the five constitutive flow law parameters that we usually have to determine using for example an inverse identification procedure as the one proposed by Dalverny \eal \cite{Dalverny-2002-ILC} from Taylor impact tests.

Analytical expressions of the three derivatives of the Johnson-Cook flow stress $\sigma^{y}$ with respect to $\varepsilon^p$, $\mdot{\varepsilon}^p$ and $T$ are given by the three following equations:
\begin{equation}
\left\{ \begin{array}{lll}
\partial \sigma^y/\partial \varepsilon^p & = & nB\varepsilon^{p^{n-1}}\left[1+C\ln\left(\frac{\mdot{\varepsilon}^p}{\mdot{\varepsilon}_0}\right)\right] \left[1-\left(\frac{T-T_0}{T_m-T_0}\right)^{m}\right]\\
\partial \sigma^y/\partial\mdot{\varepsilon}^p & = & \frac{C}{\mdot{\varepsilon}^p}\left(A+B\varepsilon^{p^{n}}\right)\left[1-\left(\frac{T-T_0}{T_m-T_0}\right)^{m}\right]\\
\partial \sigma^y/\partial T & = & \frac{-m\left(A+B\varepsilon^{p^{n}}\right)}{T-T_0}\left[1+C\ln\left(\frac{\mdot{\varepsilon}^p}{\mdot{\varepsilon}_0}\right)\right]\left(\frac{T-T_0}{T_m-T_0}\right)^{m}
\end{array}\right.
\label{eq:JCHard2}
\end{equation}

The usual approach to implement a new constitutive law in the Abaqus FEM code, as proposed by Ming \eal \cite{Ming-2018-ERV}, is to program in a VUHARD FORTRAN subroutine the evaluation of the flow stress defined by equation (\ref{eq:Johnson-Cook}) and the derivatives of the flow stress given by equation (\ref{eq:JCHard2}).

\subsection{Generation of the training and test data}
A 42CrMo4 steel, with material parameters proposed by Sattouf \eal \cite{Sattouf-2000-MIC} and reported in Table \ref{tab:Material-parameters-of-42CrMo4} was selected as the material used in this study.
\begin{table}[h]
\noindent \begin{centering}
\caption{Material properties of the 42CrMo4 steel \cite{Sattouf-2003-CDR}\label{tab:Material-parameters-of-42CrMo4}}
\par\end{centering}
\noindent \centering{}
\begin{tabular}{c|c|c|c|c|c|c}
$E$ & $\nu$ & $A$ & $B$ & $C$ & $n$ & $m$\\
(GPa) &  & (MPa) & (MPa) & &  &  \\
\hline 
$206.9$ & $0.29$ & $806$ & $614$ & $0.0089$ & $0.168$ & $1.1$\\
\hline \hline
 $\mdot{\varepsilon}_0$ & $T_0$ & $T_m$ & $\rho$ & $C_p$ & $\alpha$ & $\eta$\\
 $(s^{-1})$ & $^\circ$C & $^\circ$C & $kg/m^3$ & $J/kg^\circ$C & $10^{-6}/^\circ$C\\
\hline 
 $1$ & $20$ & $1540$ & $7830$ & $460$ & $12.3$ & $0.9$\\
\end{tabular}
\end{table}
To train and validate the ANN, we have used a Python program to generate, thanks to equations (\ref{eq:Johnson-Cook}-\ref{eq:JCHard2}) two distinct data sets:
\begin{itemize}
\item The first data set, the training one, contains $2\,520$ datapoints defined by $70$ equidistant values for $\varepsilon^p\in[0,1]$, $6$ plastic strain rates $\mdot{\varepsilon}^p\in[1, 10, 50, 500, 5\,000, 50\,000]$ and $6$ temperatures $T\in[20, 100, 200, 300, 400, 500]$.
\item The second data set, the testing set, contains $5\,000$ datapoints randomly generated within the ranges $\varepsilon^p\in[0,1]$, $\mdot{\varepsilon}^p\in[1, 50\,000]$ and $T\in[20, 500]$. This later is not used during the training phase, but only during the validation presented in section \ref{subsec:PAANN}.
\end{itemize}
Both data sets contains the values of the plastic strain $\varepsilon^p$, the plastic strain rate $\mdot{\varepsilon}^p$, the temperature $T$, the flow stress $\sigma^{y}$ computed from equation (\ref{eq:Johnson-Cook}). The second data set contains also the values of the three derivatives computed from equation (\ref{eq:JCHard2}). 
It is important to remember that the neural network proposed must allow to replace the analytical formulation of the behavior law in order to be able to carry out a numerical simulation from experimental data resulting from thermomechanical tests. During these tests, we acquire the temperature, the strain, the strain rate and the stresses. But it is impossible to acquire the derivatives of the stresses with respect to the quantities $\varepsilon^p,\mdot{\varepsilon}^p,T$. Therefore, in normal use, it is not possible to include the derivatives in the objective function for training the neural network.

\subsection{Training of the Artificial Neural Network\label{subsec:TANN}}
Training the Artificial Neural Network is finding the best set of values for all the training parameters $\w\lay{k}$, $\overrightarrow{b}\lay{k}$, $\overrightarrow{w}$ and $b$ defined in section \ref{subsec-2:NNGovEqu} in order to reduce the error of the ANN in computing the flow stress. The training set is used in this procedure and the ANN has been implemented using the Tensorflow Python library \cite{Abadi-2016-TSL}. Training procedure is based on the use of the Adaptive Moment Estimation (ADAM) \cite{Kingma-2015-AMS} optimizer. We used the two activation functions presented earlier, the $\sigmoid$ and the $\tanh$ functions, and one or two hidden layers with a variable number of neurons in the hidden layers.
All models are named after their constitution, where 3-x-1-tanh refers to a one hidden layer ANN with a $\tanh$ activation function and $x$ neurons in the hidden layer and 3-x-y-1-sig refers to a two hidden layers ANN with a $\sigmoid$ activation function in both layers and $x$ neurons in layer 1 and $y$ neurons in layer 2.
All models have been trained for the same number of iterations ($50\,000$ iterations). Training times for all models are more or less the same: around $1$ hour on a Dell XPS 13 laptop.

\subsection{Performance analysis of the Artificial Neural Network\label{subsec:PAANN}}
\begin{figure}[h] 
  \centering
  \includegraphics[width=0.8\columnwidth]{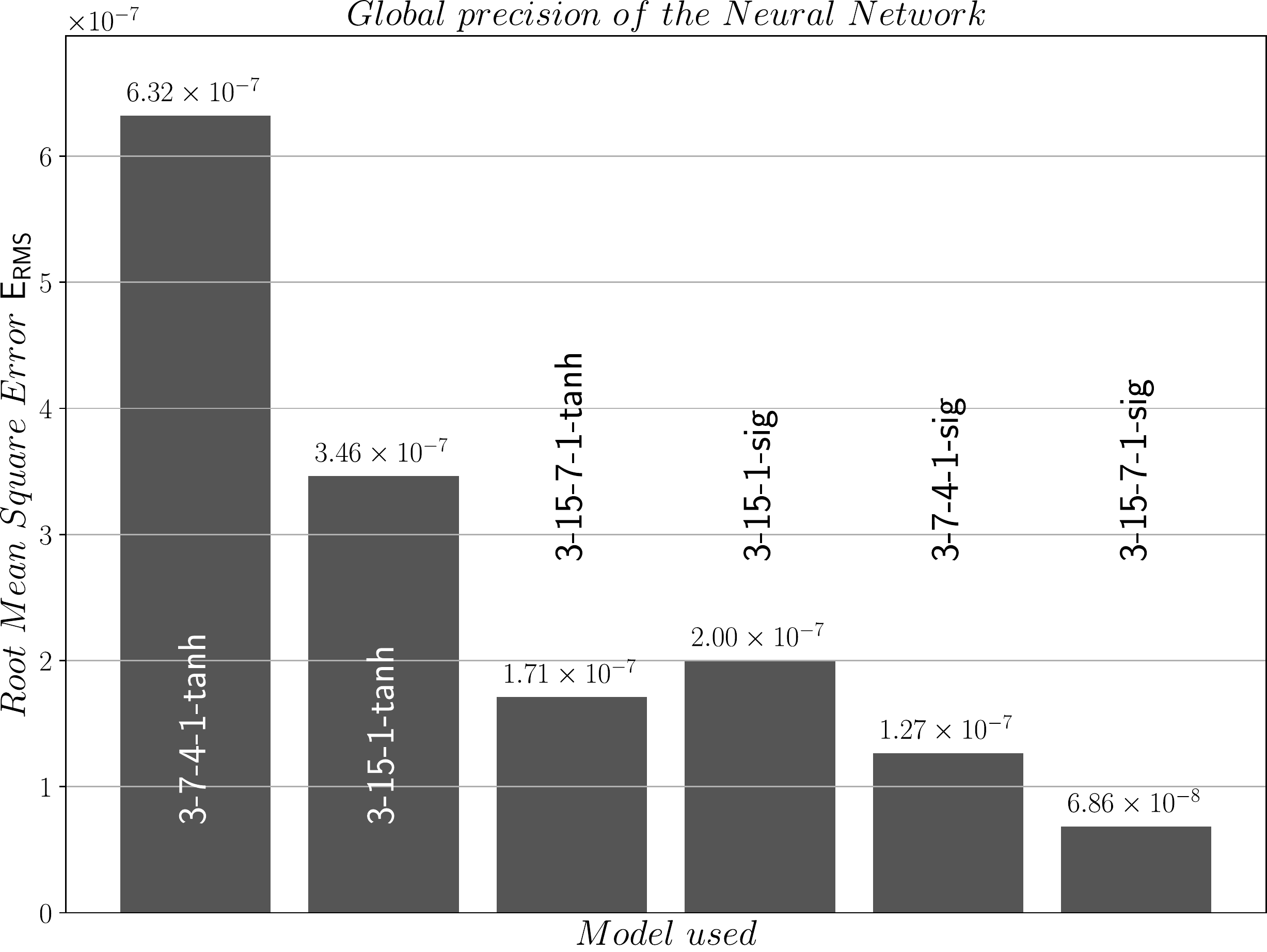}
  \caption{Comparison of precision of various ANN}
  \label{fig:ANN-precision}
\end{figure}

In order to illustrate the efficiency of the ANN, Figure \ref{fig:ANN-precision} and Table \ref{tab:ANN-report} reports some results obtained after training 6 different models ($2$ one hidden layer models and $4$ two hidden layers models).
\begin{table}[h]
\noindent \begin{centering}
\caption{Global performance analysis of the ANN during the training phase\label{tab:ANN-report}}
\par\end{centering}
\noindent \centering{}
\begin{tabular}{l|cccccc}
\multirow{2}{*}{Model} & \multirow{2}{*}{$N$} & $\ERMS$ & $\Delta \sigma$ & $\Delta (\partial \sigma/\partial \varepsilon^p)$ & $\Delta (\partial \sigma/\partial\mdot{\varepsilon}^p)$ & $\Delta (\partial \sigma/\partial T$) \\
& & $\times 10^{-7}$ & $\%$ & $\%$ & $\%$ & $\%$ \\
\hline
3-7-4-1-tanh &$65$ & $6.32$ & $0.038$ & $1.977$ & $0.792$ & $0.556$\\
3-15-1-tanh &$78$ & $3.46$ & $0.039$ & $1.506$ & $0.269$ & $0.371$\\
3-15-7-1-tanh &$180$ & $1.71$ & $0.030$ & $0.519$ & $0.380$ & $0.408$\\
3-15-1-sig &$78$ & $2.00$ & $0.030$ & $0.686$ & $0.521$ & $0.675$\\
3-7-4-1-sig &$65$ & $1.27$ & $0.024$ & $0.670$ & $0.415$ & $0.499$\\
3-15-7-1-sig &$180$ & $0.68$ & $0.011$ & $0.247$ & $0.199$ & $0.256$
\end{tabular}
\end{table}
A lot more results have been obtained in this study, but only those $6$ models are presented here to illustrate the tendencies. Figure \ref{fig:ANN-precision} shows an histogram of the average values of the $\ERMS$ defined by equation (\ref{eq:RMSE}) evaluated during the last $5\%$ of the training process giving an idea of the global convergence of the ADAM algorithm. In Table \ref{tab:ANN-report}, $N$ is the number of internal parameters of the ANN. Values $\Delta \square$ are the so-called Average Absolute Relative Error ($\AARE$) given by the following expression:
\begin{equation}
\Delta \square=\frac{1}{N} \sum_{i=1}^{N}{\left|\frac{\square_i^e -\square_i^p}{\square_i^e}\right|}
\end{equation}
where $\square_i^e$ is the analytical exact value obtained from equations (\ref{eq:Johnson-Cook}-\ref{eq:JCHard2}) and $\square_i^p$ is the ANN predicted value of the same quantity computed by the neural network from equations (\ref{eq:vonMises-ANN}) and (\ref{eq:derivatives-ANN}).

Reading the data in Table 2, the overall performance of the proposed neural networks is very good since the error on the evaluation of the flow stress is about $0.01\%$ for the best performing network while it does not exceed $0.04\%$ for all the tested networks. The evaluation of the derivatives is also very good since the error is about $0.2\%$ for the best performing network while it remains below $2.0\%$ for the $\partial \sigma/\partial \varepsilon^p$ term, $0.8\%$ for the $\partial \sigma/\partial\mdot{\varepsilon}^p$ term and $0.6\%$ for the $\partial \sigma/\partial T$ term.
Without too much surprise, we notice that globally, the error on the evaluation of the flow stress is lower with a factor of $10$ than the error on the evaluation of the derivatives. This is easily explained by the fact that the network has been trained to minimize the error on the flow stress only. Nevertheless, without ever having been optimized for computing the derivatives, we can see that the results obtained are also very good.

\begin{figure}[h] 
  \centering
  \includegraphics[width=0.8\columnwidth]{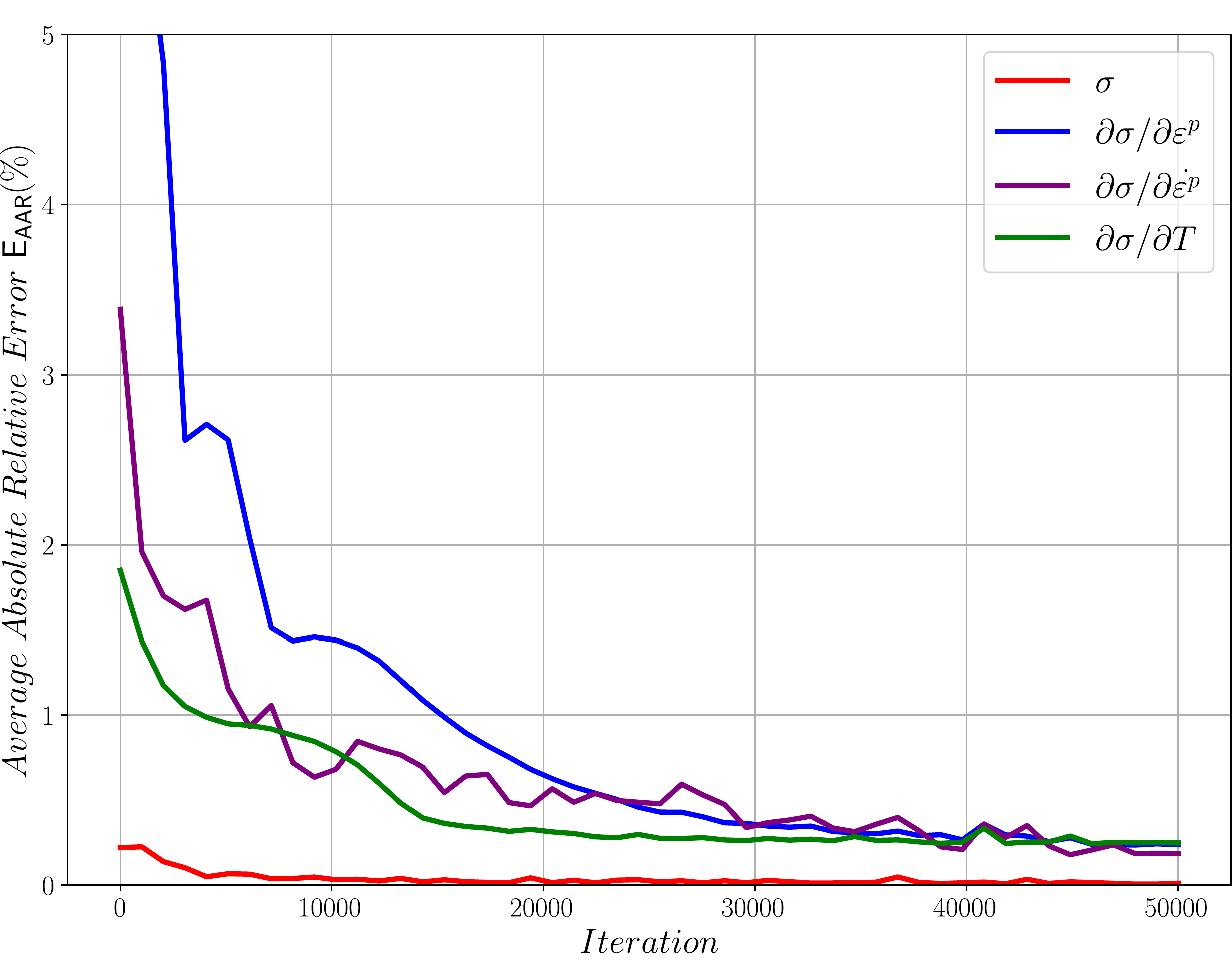}
  \caption{Convergence of the predicted values for the 3-15-7-1-sig ANN}
  \label{fig:ANN-convergence}
\end{figure}

Figure \ref{fig:ANN-convergence} shows the convergence of the 3-15-7-1-sig ANN model concerning the evaluation of the flow stress and the $3$ derivatives with the number of iterations. 
From this later, one can see that the convergence of the ANN on the evaluation of $\sigma$ is faster than convergence of the derivatives.
The previously mentioned factor of $10$ between the stress and the derivatives is clearly visible on this graph. If the error on the calculation of the constraint does not evolve after $10\,000$ iterations of the learning algorithm, the errors on the evaluation of the derivatives require a much more significant number of iterations. It is therefore necessary to continue training this type of model when the convergence criterion of the flow constraint has already been satisfied in order to allow the convergence of the derivatives. We can then be confronted with a problem of over-learning, it is therefore necessary to dimension the size of the neural network as accurately as possible with respect to the application.

\section{Implementation of the neural network in Abaqus Explicit\label{sec-4:ImpNNAbaExpl}}
\begin{figure}[h!] 
  \centering
  \includegraphics[width=0.5\columnwidth]{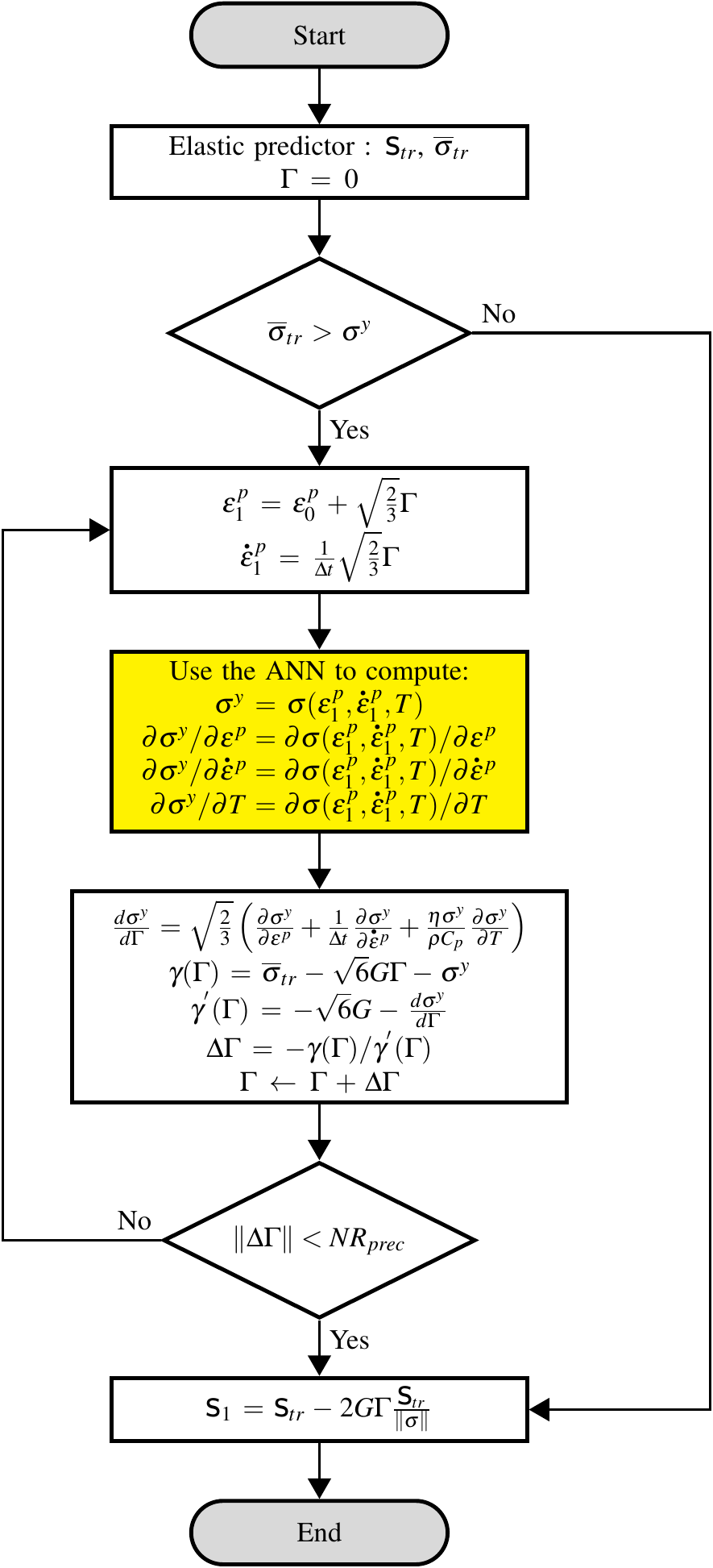}
  \caption{Flowchart of the Radial-Return Algorithm to compute the final stresses}
  \label{fig:RR-algorithm}
\end{figure}

In this section, we now present the numerical implementation of the proposed ANN constitutive flow law. Once the ANN network presented in section \ref{sec-2:ANN-Setup} has been trained as presented in section \ref{sec-3:TrainPerfEval}, it is time to use it for the flow stress and its three derivatives computing. The implementation is done by programming a VUHARD subroutine for the Abaqus Explicit finite element code similarly to the approach proposed by Jansen van Rensburg \eal \cite{JansenVanRensburg-2012-TSV}. This VUHARD subroutine is used inside of a Radial-Return algorithm, as illustrated in Figure \ref{fig:RR-algorithm}, to compute the flow stress $\sigma^y$ and its derivative $\frac{d\sigma^{y}}{d\Gamma}$ used in the expression of the two quantities $\gamma(\Gamma)$ and $\gamma^{'}(\Gamma)$, thanks to the following equation \cite{Ming-2018-ERV}:
\begin{eqnarray}
\frac{d\sigma^{y}}{d\Gamma} & = & \frac{\partial\sigma^{y}}{\partial\varepsilon^{p}}\frac{d\varepsilon^{p}}{d\Gamma}+\frac{\partial\sigma^{y}}{\partial\mdot{\varepsilon}^p}\frac{d\mdot{\varepsilon}^p}{d\Gamma}+\frac{\partial\sigma^{y}}{\partial T}\frac{dT}{d\Gamma}\nonumber \\
 & = & \sqrt{\frac{2}{3}}\left(\frac{\partial\sigma^{y}}{\partial\varepsilon^{p}}+\frac{1}{\Delta t}\frac{\partial\sigma^{y}}{\partial\mdot{\varepsilon}^p}+\frac{\eta\sigma^{y}}{\rho C_{p}}\frac{\partial\sigma^{y}}{\partial T}\right)\label{eq:derivatives}
\end{eqnarray}
where $\Gamma$ is the consistency parameter used in the Radial-Return algorithm as defined by Simo \eal \cite{Simo-1998-CI}, $\Delta t$ is the time increment, $\eta$ is the Taylor-Quinney coefficient defining the amount of plastic work converted into heat energy, $C_{p}$ is the specific heat coefficient and $\rho$ is the density of the material. So that, the ANN is used to compute the value of the flow stress $\sigma^y$ and the three derivatives of the flow stress with respect to $\varepsilon^p$, $\mdot{\varepsilon}^p$ and $T$ involved in equation (\ref{eq:derivatives}). This is illustrated in Figure  \ref{fig:RR-algorithm} where the yellow block in the center of the flowchart is where the ANN is used. Since this ANN is in the center of a CPU intensive loop involved for all integration points of the FEM model, this has to be optimized to reduce computing times as it will be presented further. More details on the Radial-Return algorithm can be found in Simo \eal \cite{Simo-1998-CI} and more details concerning the implementation of this algorithm in Abaqus Explicit can be found in Ming \eal \cite{Ming-2018-ERV}, where the same approach was used, except the fact that in this paper the flow stress and its derivatives were computed analytically.

Validation of the proposed implementation is done on several benchmark tests by comparing the results obtained using the ANN to native Johnson-Cook law (named Built-in hereafter) and the analytical VUHARD implementations proposed previously by Ming \eal \cite{Ming-2018-ERV} (named Analytical hereafter).

\subsection{Numerical implementation of the neural network\label{sec-4:NumImpl}}
The numerical implementation of the neural network is done using a VUHARD subroutine for the Abaqus Explicit code. This is a straightforward approach to implement a new constitutive flow law in this FEM code by just implementing a FORTRAN subroutine to compute the flow stress of the material $\sigma^{y}(\varepsilon^p,\mdot{\varepsilon}^p,T)$ according to equation (\ref{eq:vonMises-ANN}) and its derivatives with respect to $\varepsilon^p$, $\mdot{\varepsilon}^p$ and $T$ defined by equation (\ref{eq:derivatives-ANN}). In this approach, the main part of the Built-In constitutive law is used for time integration of the stress, for a given time increment, and the provided user subroutine is called to compute the hardening flow law.

A Python's program has been developed to extract the internal parameters of the trained neural network (the network architecture, the weights and bias values $\w$ and $\overrightarrow{b}$ of all layers,...) and write automatically the FORTRAN subroutine. In the following paragraph, we detail the implementation of a $2$ hidden layers neural network with a $\sigmoid$ activation function, so that, one of the most complex model presented in this paper.

As presented in section \ref{sec-2:ANN-Setup} the neural network is based on two main parts defined in subsections \ref{subsec-2:NNGovEqu} and \ref{subsec:DC} corresponding to the computation of the von Mises equivalent stress $\sigma$ and the $3$ derivatives $\partial \sigma/\partial \varepsilon^p$, $\partial \sigma/\partial\mdot{\varepsilon}^p$ and $\partial \sigma/\partial T$ respectively. If we want to implement a $2$ hidden layers ANN with a $\sigmoid$ activation function containing $m$ neurons in layer $1$ and $n$ neurons in layer $2$, we must use equation (\ref{eq:sigmoid}) for the computation of the stress and equation (\ref{eq:der-2-sig}) for the derivatives.
In order to optimize the numerical implementation a bit, and, as the computation of the flow stress and its derivatives share some common terms that will be stored during the computation for later use, the computation of equations (\ref{eq:sigmoid})  and (\ref{eq:der-2-sig}) is split into several sub-terms $\overrightarrow{z}^a$ to $\overrightarrow{z}^f$.
So, starting from equation (\ref{eq:preprocess}), defining the values of $\overrightarrow{x}$, one can write:
\begin{equation}
\begin{array}{ll}
z^a_i = \e{-\underset{j}{\textstyle \sum} \left(w_{ij}\lay{1} x_j\right) - b_i\lay{1}}\quad &i\in[1,m],\ j\in[1,3]\\
z^b_i = 1 + z^a_i& i\in[1,m]\\
z^c_i = \e{-\underset{j}{\textstyle \sum}\left(w_{ij}\lay{2}/ z^b_j\right) - b_i\lay{2}}& i\in[1,n],\ j\in[1,m]\label{eq:fort-derivatives-ANN-1}
\end{array}
\end{equation}
where $z^a_i$, $z^c_i$ and $z^c_i$ are three terms present in equation (\ref{eq:der-2-sig}) corresponding to $\e{-\overrightarrow{y}\lay{1}}$, $1+\e{-\overrightarrow{y}\lay{1}}$ and $\e{-\overrightarrow{y}\lay{2}}$ respectively.
Then for the computation of the derivatives, we combine the values $z^a_i$, $z^b_i$ and $z^c_i$ to compute the whole equation (\ref{eq:der-2-sig}) using:
\begin{equation}
\begin{array}{ll}
z^d_i = w_i z^c_i / (1 + z^c_i)^2\qquad & i\in[1,n]\\
z^e_i =z^a_i / z^b_i& i\in[1,m]\\
z^f_i = \underset{j}{\textstyle \sum}\left(w\lay{2}_{ji} z^d_j\right) z^e_i&i\in[1,m],\ j\in[1,n]
\end{array}
\end{equation}
From those definitions, one can then write the output of the neural network using the following expression:
\begin{equation}
s = \underset{i}{\textstyle \sum}\left(w_i / (1+z^c_i)\right) + b\qquad  i\in[1,n]\label{eq:fort-vonMises-ANN}
\end{equation}
And, the three derivatives $s'_i$ are obtained from:
\begin{equation}
s'_i = \underset{j}{\textstyle \sum}\left(w\lay{1}_{ji} z^f_j\right)\qquad i\in[1,3],\ j\in[1,m]\label{eq:fort-derivatives-ANN}
\end{equation}

\begin{figure}[h]
 \centering
 \includegraphics[width=0.95\columnwidth]{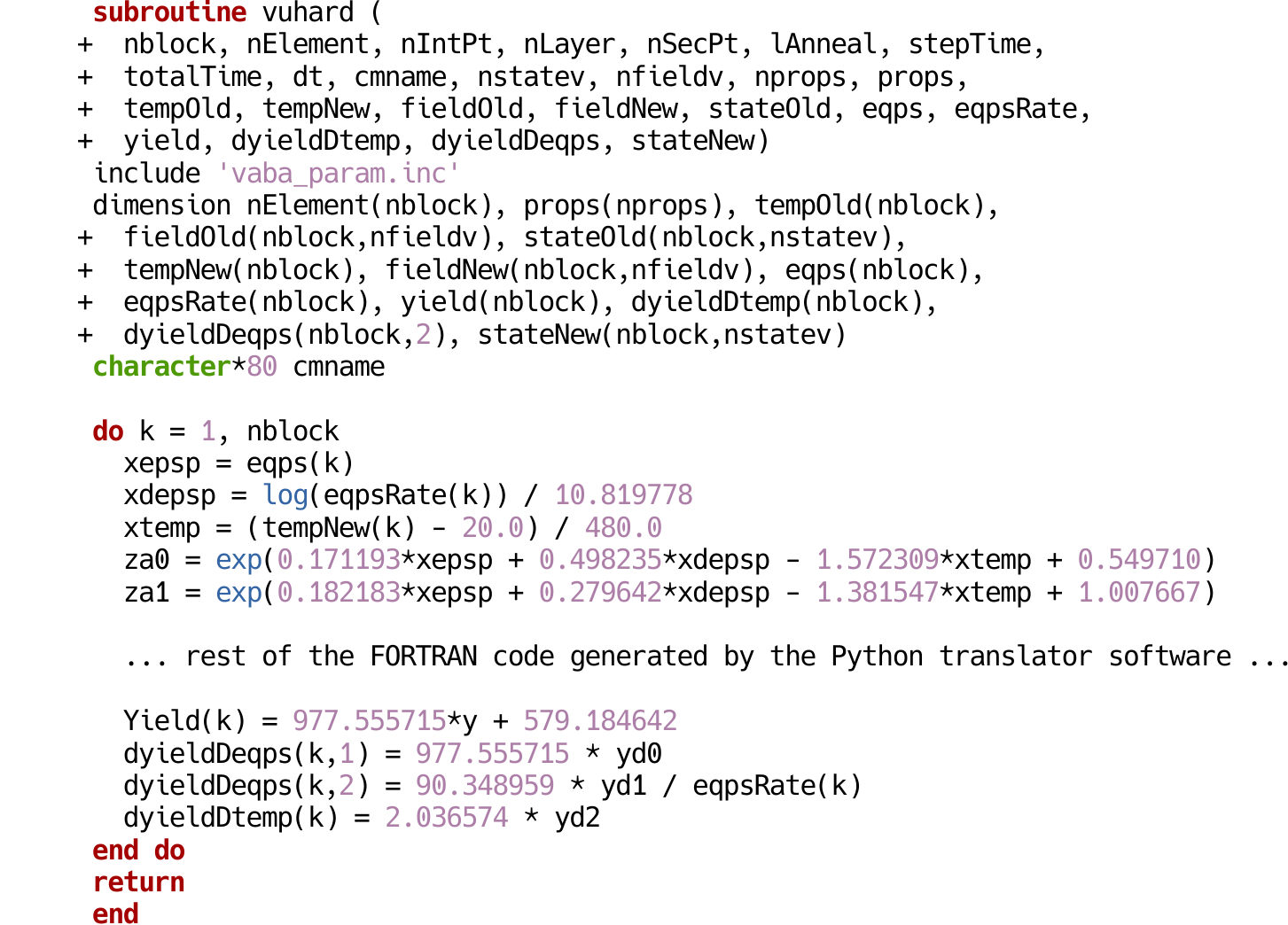}
 \caption{Partial VUHARD FORTRAN subroutine to compute the flow stress and its derivatives using the ANN for the 3-15-7-1-sig model
 \label{fig:FortranSubroutine}}
\end{figure}

Finally, equations (\ref{eq:vonMises-ANN}) and (\ref{eq:derivatives-ANN}) are used to compute the von Mises equivalent stress $\sigma$ and the $3$ derivatives $\partial \sigma/\partial \varepsilon^p$, $\partial \sigma/\partial\mdot{\varepsilon}^p$ and $\partial \sigma/\partial T$ of the neural network from $s$ and $\overrightarrow{s}'$ computed from equations (\ref{eq:fort-vonMises-ANN}) and (\ref{eq:fort-derivatives-ANN}). As it is a straightforward implementation from equation (\ref{eq:fort-derivatives-ANN-1}) to (\ref{eq:fort-derivatives-ANN}) the Python interface uses loops to explicitly write all the matrices products in a FORTRAN subroutine as illustrated in Figure \ref{fig:FortranSubroutine}. This later only reports a small part of the whole FORTRAN code to illustrate how it is implemented. The reader interested in the details of this implementation can refer to the Software Heritage archive website \cite{Code} giving access to the source code of this work.

The VUHARD subroutine is compiled using the GNU gfortran 9.3.0 and linked to the main Abaqus Explicit executable.
All benchmarks tests have been solved using Abaqus Explicit 2021 on a Dell XPS 13 laptop running Ubuntu 20.04 64~bits with
16~GiB of Ram and one 4 core i7-10510U Intel Processor. All computations have been done using the double precision option of Abaqus, with parallel threads execution on two cores. 
In order to reduce the number of models presented hereafter, only the two last models presented in Table \ref{tab:ANN-report} are selected for the subsequent benchmark simulations.

\subsection{Necking of a circular bar benchmark test\label{subsec:Necking}}

\begin{figure}[h]
 \centering
 \includegraphics[width=0.8\columnwidth]{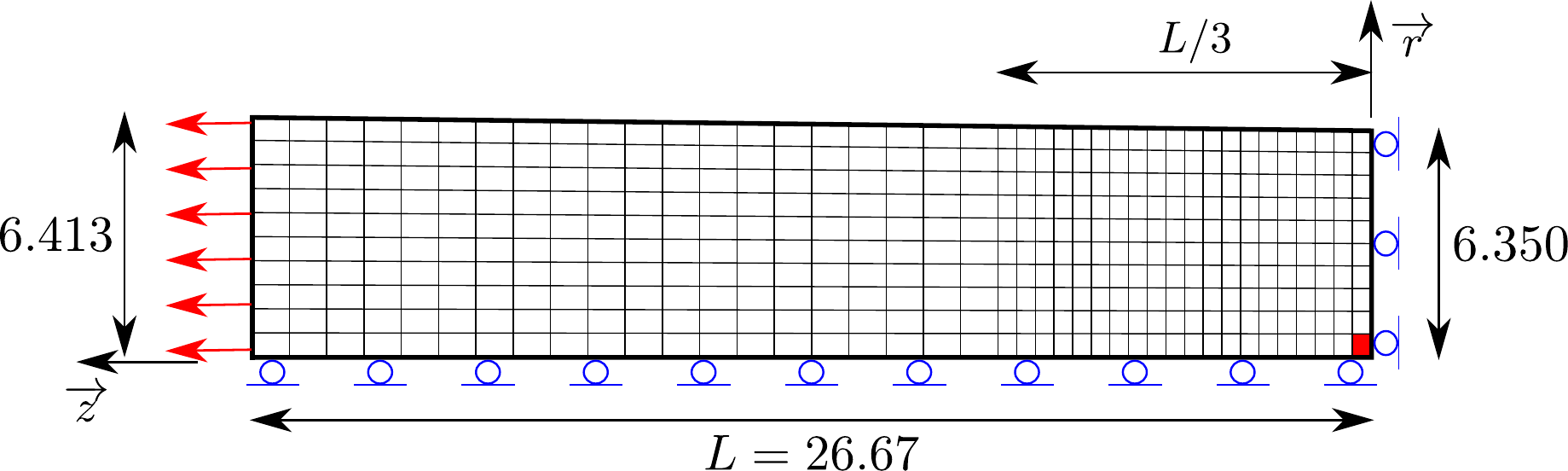}
 \caption{Numerical model for the necking of a circular bar
 \label{fig:BarNeckingModel}}
\end{figure}

The necking of a circular bar test, already presented by Ponthot \eal \cite{Ponthot-2002-USU}, is useful to evaluate the performance of non-linear constitutive laws. An axisymmetric quarter model of the specimen, already presented in Ming \eal \cite{Ming-2018-ERV}, is used, where dimensions of the specimen are reported in Figure \ref{fig:BarNeckingModel}. We imposed a total displacement of $7~\text{mm}$ along the $\overrightarrow{z}$ axis on the left side of the specimen while the radial displacement of the same edge is supposed to remain zero. On the opposite side, the axial displacement is restrained while the radial displacement is free. The mesh consists of $400$ CAX4RT elements (4-node bilinear displacement and temperature, reduced integration with hourglass control element) with a refined zone of $200$ elements on the right side on $1/3$ of the total length. The FEM model is a coupled temperature-displacement explicit model (this is a coupled thermal-stress analysis where the heat transfer and mechanical solutions are obtained simultaneously by an explicit coupling), and the total simulation time is set to $t=0.01~\text{s}$. 

\begin{figure}[h]
 \centering
 \includegraphics[width=0.8\columnwidth]{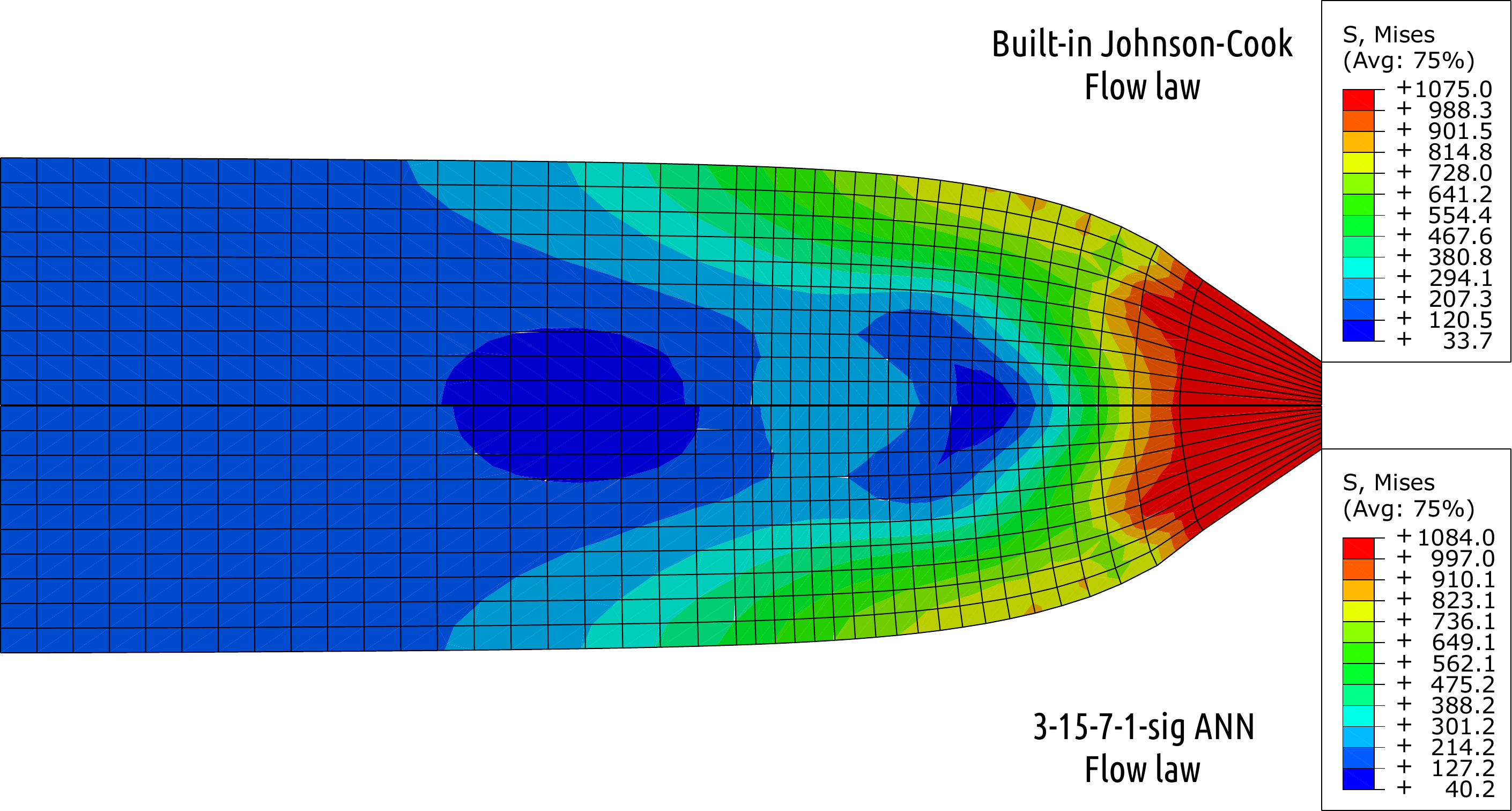}
 \caption{Von Mises stress $\overline{\sigma}$ contourplot for the necking of a circular bar for an elongation of $7~\text{mm}$ (top side is the Built-in flow law and bottom side is the ANN 3-15-7-1-sig flow law).
 \label{fig:misesCPBarNecking}}
\end{figure}

\begin{figure}[h] 
  \centering
  \includegraphics[width=0.8\columnwidth]{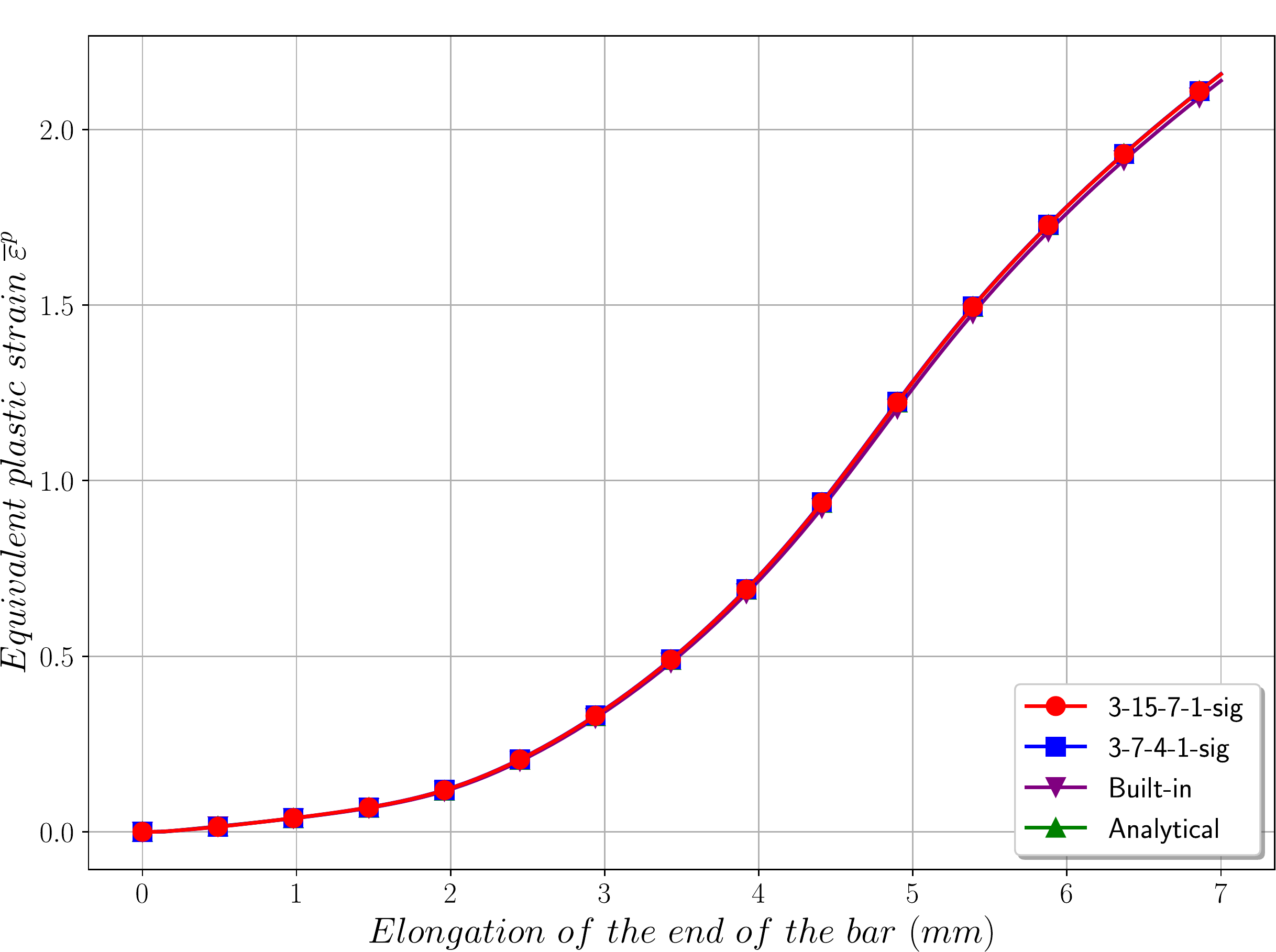}
  \caption{Equivalent plastic strain $\overline{\varepsilon}^p$ \versus displacement for the necking of a circular bar}
  \label{fig:strainBarNecking}
\end{figure}

\begin{figure}[h]
  \centering
  \includegraphics[width=0.8\columnwidth]{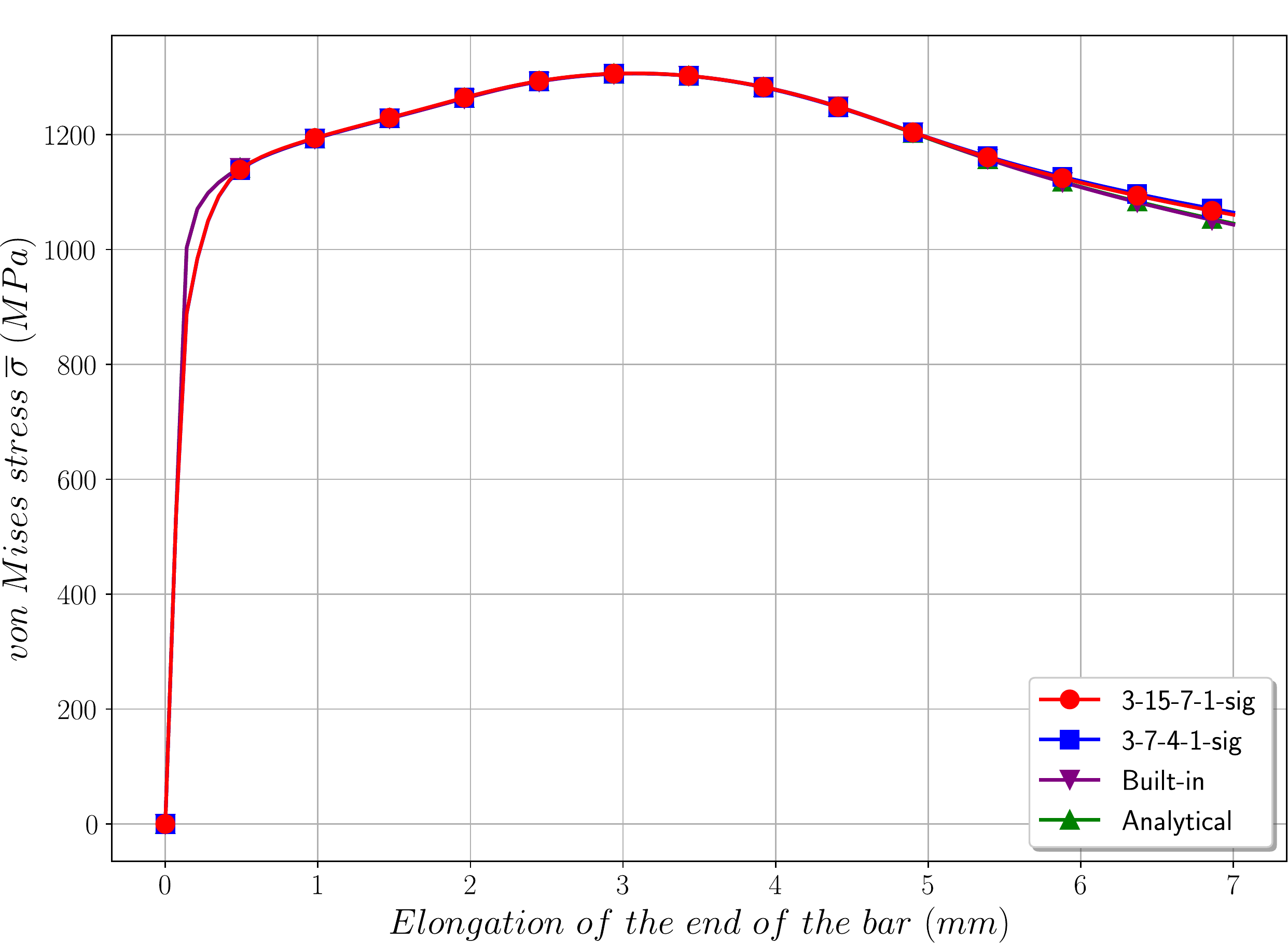}
  \caption{Von Mises stress $\overline{\sigma}$ \versus displacement for the necking of a circular bar}
  \label{fig:misesBarNecking}
\end{figure}

Figure \ref{fig:misesCPBarNecking} shows the von Mises stress contourplot $\overline{\sigma}$ of the deformed bar for two different models: the Built-In model (top side) and the ANN 3-15-7-1-sig model (bottom side). There is a very little difference between both models in terms of spatial distribution of stress and maximum value. The maximum stress is located into the center of the bar, so we have chosen to plot in Figures \ref{fig:strainBarNecking} and \ref{fig:misesBarNecking} the evolution of the equivalent plastic strain $\overline{\varepsilon}^p$ and the von Mises stress $\overline{\sigma}$ for the central element of the specimen (the red element in the bottom right corner in Figure \ref{fig:BarNeckingModel}). As reported in Figures \ref{fig:strainBarNecking} and \ref{fig:misesBarNecking}, the Built-In model, the Analytical model and both versions of the ANN model give almost the same results, except when the elongation is greater than $6~\text{mm}$ where the von Mises stress differs between the ANN models and the Built-in and Analytical ones as we will see further.

This is also confirmed in Table \ref{tab:BarNeckingTable} reporting a comparison of the four models for two values of the displacement ($3.5~\text{mm}$ and $7~\text{mm}$, named $mid$ and $end$ respectively). From this later, we can see that the equivalent plastic strain $\overline{\varepsilon}^p$, the von Mises stress $\overline{\sigma}$ and the temperature $T$ obtained with both ANN models are very close to the ones obtained by the Built-in and the Analytical models for $mid$ displacement, while they differ a little for $end$ displacement. This allows us to validate the proposed approach.

One interesting result concerns the values of the equivalent plastic strain $\overline{\varepsilon}^p$ and the temperatures $T$ reported in Table \ref{tab:BarNeckingTable} for $end$ displacement. It can be seen that, for $end$ displacement, the values of the plastic strain reported in Table \ref{tab:BarNeckingTable} are around $2.1$ while the range of plastic strain used for training the model is $[0,1]$. Furthermore, the maximum value of the temperature is around $587^\circ$C while the training range is $[20,500]$. The proposed model is therefore able to extrapolate with a good accuracy some data out of the training range.
It is obvious from Table \ref{tab:BarNeckingTable} and Figures \ref{fig:strainBarNecking} and \ref{fig:misesBarNecking} that all models give very closed results when parameters are within the training range, while the ANN models differ a little when parameters are far out of the training range (here, with an elongation greater than $6~\text{mm}$ so that $\overline{\varepsilon}^p>1.7$). This shows that our ANN model is able, to a certain extent, and with the usual precautions insofar as neural networks are generally relatively faithful for interpolation but less so for extrapolation, to generalize a behavior out of the training range. If we want to reduce the gap between predicted and real values out of the training range, we have to enlarge the training ranges of the input variables if data is available.

Table \ref{tab:BarNeckingTable} reports also the computing times and total number of increments computed after running the same model $10$ times. One can see that computation time of the 3-7-4-1-sig model is equivalent to the one of the Built-in model and lower to the one of the Analytical model while there is an increase for the most complex ANN model. In the proposed approach, the ANN is used to compute the flow stress of the material within a Radial Return algorithm. It replaces the evaluation of the flow stress and its derivatives based on some analytical expressions. VUHARD implementation breaks the Built-in natural optimized algorithm used to compute the stresses by some FORTRAN subroutine call, data transfer,... leading to increase of CPU time (see comparison between Built-in and Analytical CPU times). This is why we cannot have a reduction of computational time with regard to the Built-in model. We therefore have to compare CPU performance of ANN models with the Analytical one. Since the Johnson-Cook analytical behavior law is quite simple, the most complex ANN model is slower than the analytical one. But, in case of more complex behavior law, such as some Modified Johnson-Cook laws proposed by Zhou et al \cite{Zhou-2020-RSC}, this tendency will be reversed since the analytical evaluation of the derivatives needed for the radial return algorithm to work becomes quite CPU intensive. We can also conclude from the  results presented in Table \ref{tab:BarNeckingTable} that the 3-7-4-1-sig model is sufficient to obtain valuable results with comparable CPU times with regard to the Built-in constitutive law.

\begin{table}[h]
\noindent \begin{centering}
\caption{Comparison of results for the necking of a circular bar benchmark for a displacement of $3.5~\text{mm}$ (mid) and $7~\text{mm}$ (end)\label{tab:BarNeckingTable}}
\par\end{centering}
\noindent \centering{}
\begin{tabular}{l|cccccccc}
\multirow{2}{*}{Model} & \multirow{2}{*}{Incr.} & Time &  \multirow{2}{*}{$\overline{\varepsilon}^p_{mid}$} & $\overline{\sigma}_{mid}$ & $T_{mid}$ &\multirow{2}{*}{$\overline{\varepsilon}^p_{end}$} & $\overline{\sigma}_{end}$ & $T_{end}$\\
 &  & (s) &  & (MPa) & ($^\circ$C) &  & (MPa) & ($^\circ$C)\\
\hline 
3-7-4-1-sig & $191\,768$ & $29.98$ & $0.51$ & $1293.81$ & $182.01$ & $2.16$ & $1064.43$ & $587.78$\\
3-15-7-1-sig & $194\,432$ & $38.54$ & $0.51$ & $1293.97$ & $181.52$ & $2.03$ & $1060.50$ & $587.29$\\
Analytical & $200\,145$ & $35.50$ & $0.51$ & $1293.59$ & $182.47$ & $2.16$ & $1045.75$ & $585.85$\\
Built-In & $199\,474$ & $28.71$ & $0.51$ & $1293.76$ & $180.36$ & $2.14$ & $1043.19$ & $587.66$\\
\end{tabular}
\end{table}

\subsection{Taylor impact benchmark test\label{subsec:Taylor}}

\begin{figure}[h]
 \centering
 \includegraphics[width=0.8\columnwidth]{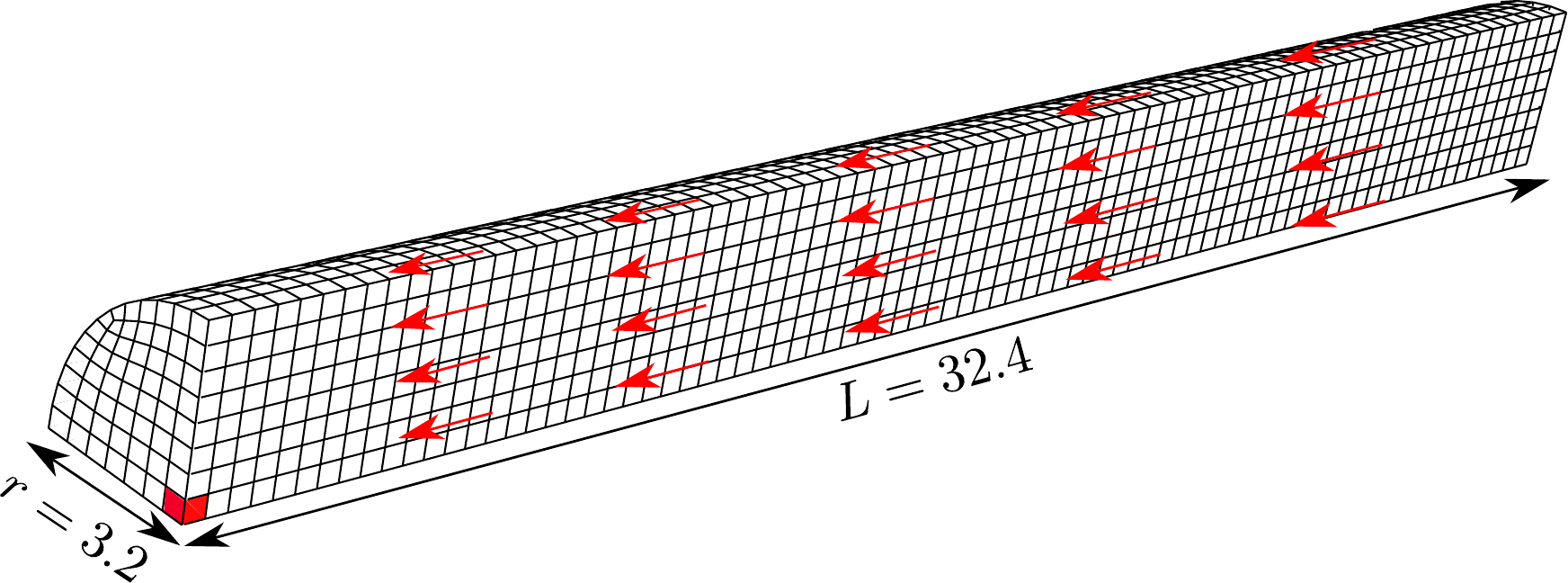}
 \caption{Numerical model for the Taylor impact test
 \label{fig:TaylorModel}}
\end{figure}

The performance of the proposed ANN subroutine will now be validated under high deformation rate with the simulation of the Taylor impact test \cite{Taylor-1946-TMH} where a cylindrical specimen is launched to impact a rigid target with a prescribed initial velocity $V_{c}=287~\text{m/s}$. The height of the cylinder is $32.4~\text{mm}$ and the radius is $3.2~\text{mm}$ as reported in Figure \ref{fig:TaylorModel}. The axial displacement is restrained on the left side of the specimen while the radial displacement is free (to figure a perfect contact without friction of the projectile onto the target). A 3D quarter model of the Taylor cylindrical specimen is meshed with $4\,455$ C3D8T elements (8-node trilinear displacement and temperature). The total simulation time for the Taylor impact test is $t=80\,\mu \text{s}$.

\begin{figure}[h]
 \centering
 \includegraphics[width=0.8\columnwidth]{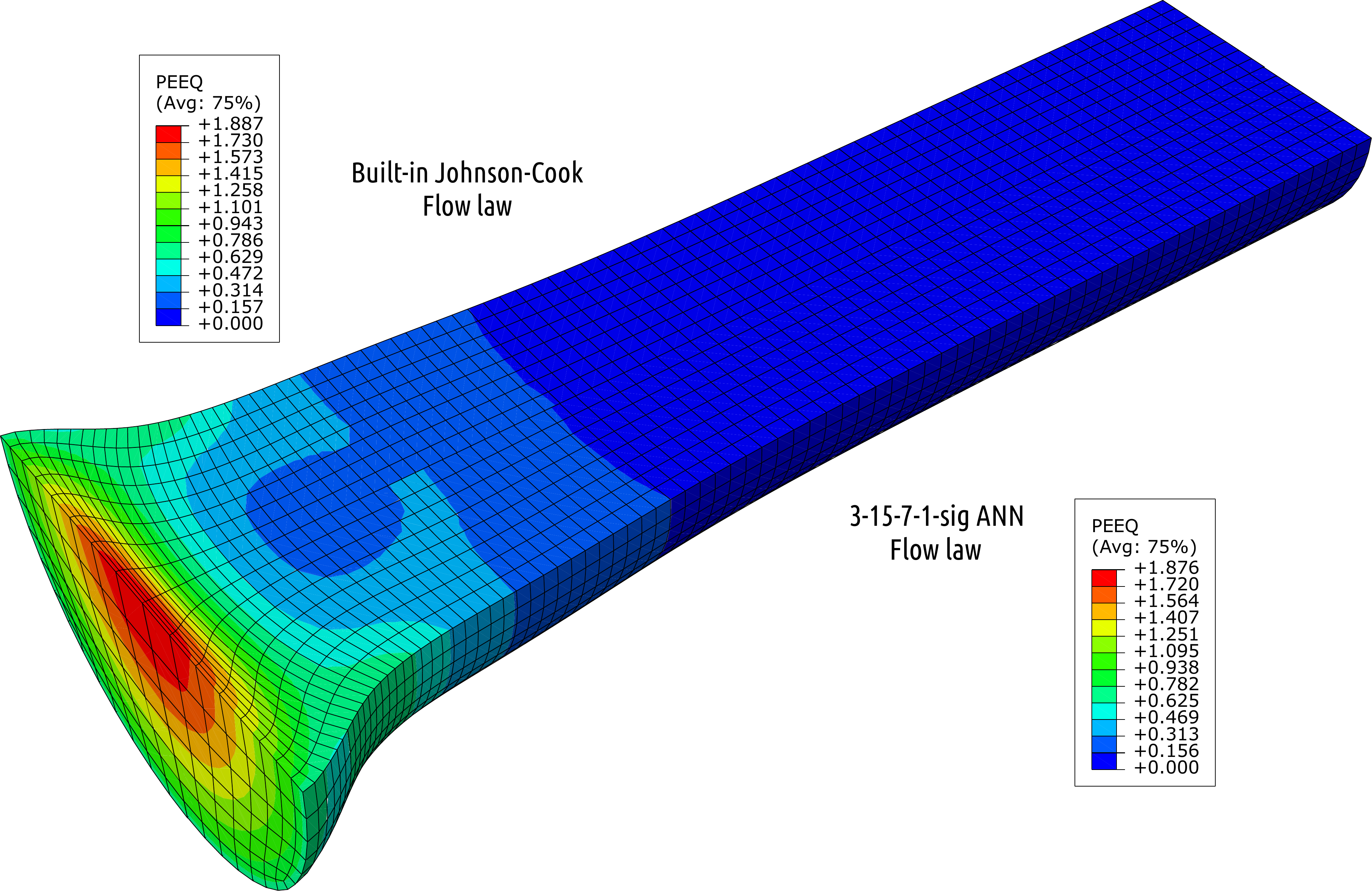}
 \caption{Equivalent plastic strain  $\overline{\varepsilon}^p$ contourplot for the Taylor impact test (upper side is the Built-in model and bottom side is the ANN 3-15-7-1-sig model)
 \label{fig:strainTaylor}}
\end{figure}

Figure \ref{fig:strainTaylor} shows the equivalent plastic strain contourplot of the deformed rod for two models: the Built-In model (upper side of the specimen) and the ANN-3-15-7-1-sig model (bottom side of the specimen). The distributions of
the equivalent plastic strain are almost the same for both models.
The maximum equivalent plastic strain $\overline{\varepsilon}^{p}$ is located into the center element of the model (the red element in Figure \ref{fig:TaylorModel}) and the models give quite the same value as reported in Table \ref{tab:TaylorTable} for $\overline{\varepsilon}^{p}$, $T$ and the final dimensions of the specimen $L_{f}$ (final length) and $D_{f}$ (final diameter of the impacting face). 
Concerning the simulation times reported in Table \ref{tab:TaylorTable}, the same trends as those presented in section \ref{subsec:Necking} are noticeable in this test case. Again, the comparison of the numerical results validates the proposed approach and shows a very good level of correlation of the results.

\begin{table}[h]
\noindent \begin{centering}
\caption{Comparison of results for the 3D Taylor impact test\label{tab:TaylorTable}}
\par\end{centering}
\noindent \centering{}
\begin{tabular}{l|cccccc}
\multirow{2}{*}{Model} & \multirow{2}{*}{Incr.} & Time & $L_{f}$ & $D_{f}$ & $T$ &\multirow{2}{*}{$\overline{\varepsilon}^p$} \\
 &  & (s) & (mm) & (mm) & $(^\circ$C)& \\
\hline 
3-7-4-1-sig & $6\,344$ & $63.08$ & $26.52$ & $11.18$ & $584.28$ & $1.83$ \\
3-15-7-1-sig & $6\,239$ & $90.32$ & $26.52$ & $11.17$ & $584.47$ & $1.83$ \\
Analytical & $6\,419$ & $71.71$ & $26.53$ & $11.19$ & $585.64$ & $1.84$ \\
Built-In & $6\,570$ & $52.82$ & $26.54$ & $11.21$ & $588.66$ & $1.84$ \\
\end{tabular}
\end{table}

\section{Summary and conclusions}
In this paper, an Artificial Neural Network based framework has been proposed to model the non-linear flow law  $\sigma^{y}(\varepsilon^p,\mdot{\varepsilon}^p,T)$ with its application to a 42CrMo4 steel and a constitutive behavior of type Johnson-Cook.
The general architecture of the multilayer perceptron neural network was presented with a focus on the evaluation of the derivatives of the flow stress with respect to the plastic strain, the plastic strain rate and the temperature necessary to implement a VUHARD user routine in the finite element code Abaqus Explicit, without these quantities having been learned by the network according to the classical supervised training scheme. The evaluation of these derivatives for $1$ or $2$ hidden layers and $2$ types of activation function has been presented in detail as well as the comparison of the accuracy of this evaluation with a reference solution based on the use of the Johnson-Cook flow law. The results obtained showed an excellent ability to evaluate the flow stress and a very good ability to evaluate the derivatives by the neural network.
After numerical implementation of the neural network in the Abaqus code, the test cases used showed the good behavior of the proposed approach in the context of the numerical simulation of the necking of a circular bar and a Taylor impact test.

\bibliographystyle{elsarticle-num}
\addcontentsline{toc}{section}{\refname}
\bibliography{bibtex}

\end{document}